\documentstyle{ioplppt}
\begin{document}
\jl{1}

\title{Polymer Adsorption on Fractal Walls}

\author{G Giugliarelli\dag, A L Stella\ddag \S}

\address{\dag INFM--Unit\`a di Padova and Dipartimento di Fisica,
Universit\`a di Udine, I--33100 Udine, Italy}

\address{\ddag INFM--Dipartimento di Fisica and Sezione INFN,
Universit\`a di Padova, I--35131 Padova, Italy}

\address{\S The Abdus Salam I. C. T. P, Trieste, I--34014 Italy}


\begin{abstract}
Polymer adsorption on fractally rough walls of varying
dimensionality is studied by renormalization group methods on
hierarchical lattices. Exact results are obtained for
deterministic walls. The adsorption transition is found continuous
for low dimension $d_w$ of the adsorbing wall and the
corresponding crossover exponent $\phi$ monotonically increases
with $d_w$, eventually overcoming previously conjectured bounds.
For $d_w$ exceeding a threshold value $d_w^*$, $\phi$ becomes 1
and the transition turns first--order. $d_w^*>d_{saw}$, the
fractal dimension of the polymer in the bulk. An accurate
numerical approach to the same problem with random walls gives
evidence of the same scenario.
\end{abstract}

\maketitle

\section{Introduction}

The adsorption on an attracting impenetrable wall is perhaps the
most elementary transition involving a single interacting polymer
in solution \cite{DBL}. High dilution in a good solvent is the
realistic condition for which this problem can be directly
relevant. The fundamental character and the obvious relation with
more complex applications, like colloid stabilization or surface
protection \cite{N}, attracted on polymer adsorption a great deal
of attention in recent years, and much information is presently
available on this problem. It is now well understood that this
transition can be interpreted as a surface critical phenomenon
\cite{DBL,V}: at the adsorption temperature $T_a$ the
conformational statistics of the polymer shows a multicritical
behavior with peculiar geometric features and with crossovers to
the high-$T$ desorbed and the low-$T$ adsorbed regimes. For a
chain with $N$ monomers at $T_a$ the average number of adsorbed
monomers, $\langle M\rangle$, scales as $\langle M\rangle\propto
N^{\phi}$, where $\phi$ ($0<\phi<1$) is the crossover exponent. In
the high- and low-$T$ regimes, $\langle M\rangle\propto N^0$ and
$\langle M\rangle\propto N$, respectively. $\phi$ is known exactly
in $2D$ for a polymer in both good \cite{BEG} and theta \cite{VSS}
solvents, and in $3D$ in theta solvent, in which case logarithmic
corrections are present \cite{DE}. Further exact results have been
obtained for models defined on fractal lattices, like Sierpinski
gaskets \cite{BV,KS,ZMS}, which are by now recognized as an
important context in which to test theoretical ideas concerning
polymer statistics.

Most explicit results obtained so far on polymer adsorption refer
to cases in which the wall is smooth and flat. In this paper we
address the adsorption transition on a fractal substrate. This
problem has applicative interest. Indeed, in many processes
involving polymers, highly corrugated, irregular walls may be
present. In addition there are interesting theoretical
implications. A polymer in good solvent is known to possess a
self--similar stochastic geometry, with a well defined fractal
dimension $d_{saw}$. Once such a polymer is put in contact with a
fractal wall, a competition between the two geometries arises.
This holds in particular for cases in $2D$ or in hierarchical
lattices, for which the wall is topologically one--dimensional
like the polymer. As we will see, the above competition leads to a
modification of the universal properties of the adsorption
transition, or, in more extreme situations, to a drastic and quite
unusual suppression of its continuous character. The parameter
triggering such modifications is the fractal dimension $d_w$ of
the wall. This scenario has also analogies with another situation
in which competition between two similar scaling geometries has
been studied recently. A fluctuating interface between two
coexisting phases is self--affine \cite{DL}. If it is put in
contact with a rough wall of similar geometry, depinning from the
wall turns from continuous to first--order as soon as the
roughness exponent of the wall exceeds the anisotropy index of
interface fluctuations in the bulk \cite{GS,SS}.

An approach to our problem on Euclidean lattice would meet very
serious difficulties. Once assigned a given profile to a fractal
wall exerting short range attraction on a self--avoiding chain
(SAW), exact enumerations would handle too short chains, unable to
feel the fractal corrugations of the wall on a sufficiently wide
range of length scales. On the other hand, Monte Carlo simulations
meet the serious obstacle that, in the low-$T$ region, sampling
over polymer configurations becomes very problematic, due to the
highly irregular wall, with valleys and hills at all scales. For
these reasons adsorption on fractal walls is certainly one of
those phenomena for which the study on simplified, hierarchical
lattices is at present the only realistic way to gain an at least
qualitative understanding. Two recent works studied adsorption on
fractal boundaries of SAW's within fractal lattices
\cite{MMZ,MZM}. Most emphasis there was put on the existence of
violations of bounds suggested in previous work \cite{BV} for
$\phi$. It was also realized that fractal lattices with peculiar
connectivities at the borders could give rise to an interesting
dependence of $\phi$ on the interaction parameters. However, such
nonuniversality, met also with flat walls, is specific of the
lattices considered, which do not mimic generic situations.
Indeed, it is natural to expect universal scaling at the
adsorption transition for a given (universal) bulk criticality and
a given boundary geometry (no matter whether flat or rough).
Changes of $\phi$ should be expected upon varying $d_w$. This is
indeed the dimension pertaining to the surface critical phenomenon
to which adsorption amounts.

Here we study adsorption in three hierarchical lattices leading to
renormalization group (RG) recursions of increasing complexity
and, supposedly, to results of increasing qualitative value with
reference to realistic situations on $2D$ Euclidean lattice.

\section{Models and Results}

\begin{figure}
\vspace{6.0cm}
\includegraphics{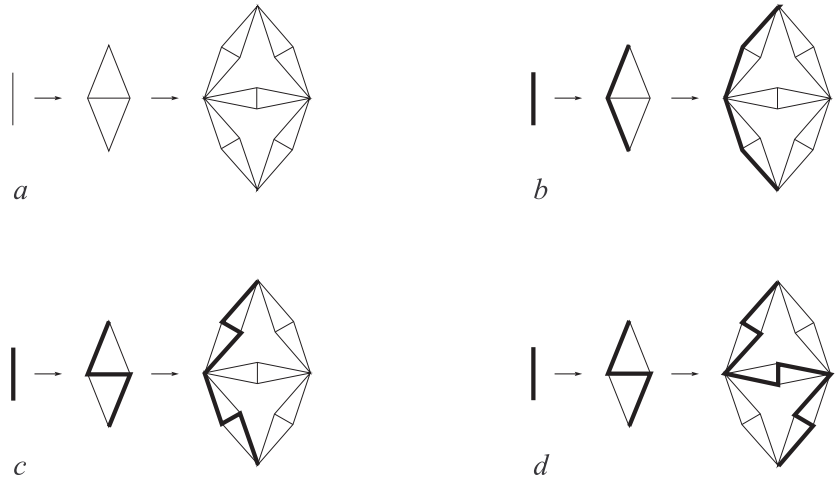}
\caption{\label{Fig:LA} $a$ Construction rule of ${\cal L}_A$. The
length of the diagonal step is different from the other lattice
steps only for drawing convenience. $b$--$d$ Construction rules of
deterministic fractal walls (heavy lines) in ${\cal L}_A$. The
allowed polymer configurations develop to the right of the walls.}
\end{figure}

Let us consider first the lattice ${\cal L}_A$, whose construction
rule is sketched in \Fref{Fig:LA}a. Measuring the lattice ``linear
size'' in terms of the number of steps of the shortest path
between top and bottom vertices, at any application of the
construction rule, this size and the total number of lattice bonds
are multiplied by factors 2 and 5, respectively. Thus, ${\cal
L}_A$ has a fractal dimension $d_L=\ln 5/\ln 2=2.322\ldots$ At any
level, $n$, in the construction of ${\cal L}_A$, allowed polymer
configurations correspond to SAW's between the top and bottom
vertices. In a grand canonical formulation, to each step is
associated a monomer fugacity $\omega= \exp({\mu}/T)$. An
attracting impenetrable wall is modeled as a particular SAW which
can not be trespassed by the polymer. The polymer interacts with
the wall through an attractive contact potential $-\epsilon$.
Thus, SAW steps on the wall acquire an extra fugacity
$k=\exp(\epsilon/T)$ ($k\geq 1$).

In the absence of wall, through the $n=0$ lattice there is a
unique walk of unit length and the restricted grand partition
function for SAW's joining top to bottom is simply $B_0=\omega$.
At level $n=1$ there are two pairs of SAW's of lengths 2 and 3,
respectively. The corresponding partition function is
$B_1=2\omega^2+2\omega^3= 2B_0^2+2B_0^3\equiv {\cal B}(B_0)$. If
we denote by $B_n$ the SAW partition at level $n$, we can write
$B_{n+1}\equiv{\cal B}(B_n)$ and $\cal B$ can be seen as a
generating function of the bulk partition function. The recursion
for $B_n$ has a repulsive fixed point at
$B^*=(\sqrt{3}-1)/2=0.366\ldots$ which corresponds to the bulk
critical point of SAW's. Thus $\omega_c=B^*$ is the SAW critical
fugacity. For an $n$-th level lattice the average number of SAW
steps is given by $\langle N_B\rangle_n={{\omega}\over{B_n}}\
{{\partial B_n}\over{\partial \omega}}$. At $\omega=\omega_c$ we
have $\langle N_B\rangle_n=\lambda_B^n$ with
$\lambda_B=\left.{{d{\cal B}(B)}\over
{dB}}\right|_{B^*}=2.268\ldots$ Therefore, taking into account
that the lattice size is $L_n=2^n$, we conclude that critical
SAW's are fractal with dimension $d_{saw}=\ln\lambda_B/\ln
2=1.181\ldots$

Now let us consider a wall through ${\cal L}_A$. In Figure 1b and
1c we sketch two examples of deterministic rules by which wall
geometries with opposite features can be realized. Iteration of
the rule sketched in Figure 1b produces a wall whose length is
$L_n$. Thus, this wall is characterized by a dimension $d_w=1$ and
we regard it as flat. On the contrary, the rule in Figure 1c
produces a fractal wall with dimension $d_w=\ln 3/\ln
2=1.585\ldots$ which is also the highest realizable in ${\cal
L}_A$. Walls with intermediate dimensions can be obtained by using
either deterministic, or random sequences of the two rules above
at progressing levels of lattice construction. As an example, let
us consider a case in which the wall is realized by means of rule
1c and 1b for odd and even $n$, respectively. The resulting wall
has a dimension $d_w=\ln 6/\ln 4=1.292\ldots$ and is sketched in
Figure 1d. For $n=1$ the SAW partition function in the presence of
the wall is given by
\begin{equation}
\fl
X_1=X_0^3+X_0B_0\equiv {\cal X}_2(B_0,X_0)
\label{Eq:LArec1}
\end{equation}
where $X_0=k\omega$ is the partition of the unique SAW in the
$n=0$ lattice. At $n=2$ we have
\begin{equation}
\fl X_2=X_1^2+2X_1B_1^2+B_1^2\equiv {\cal X}_1(B_1,X_1).
\label{Eq:LArec2}
\end{equation}
${\cal X}_1$ and ${\cal X}_2$ are now the generating functions of
the SAW partitions corresponding to rules 1b and 1c, respectively.
At any other construction level, the form of the recursions is the
same as in \Eref{Eq:LArec1} or \Eref{Eq:LArec2} for $n$ even or
odd, respectively. Focusing attention on even $n$, if $X_n$
denotes the SAW partition at level $n$, we have
\begin{equation}
\fl X_{n+2}={\cal X}_2({\cal B}(B_n),{\cal X}_1(B_n,X_n)).
\label{Eq:LAevenrec}
\end{equation}

For $B=\omega_c$ the bulk recursion is at its repulsive fixed
point, while \eref{Eq:LAevenrec} has an attractive fixed point at
$X=0.1536\ldots$ and a repulsive one at $X\equiv x_c\equiv
0.7249\ldots$ \cite{MATHEMATICA} Another attractive fixed point is
at $X=\infty$. While the first fixed point controls the SAW
ordinary desorbed regime, the second one corresponds to the
adsorption critical point which is then located at a wall
attraction $k_c=x_c/\omega_c=1.9806\ldots$ The critical exponents
can be obtained by linearization of the RG flow around the fixed
point $(\omega_c,x_c)$. In doing this, together with the recursion
\eref{Eq:LAevenrec} we have to consider also $B_{n+2}={\cal
B}({\cal B}(B_n))$ and the matrix
\begin{equation}
\fl
{\bf R}\equiv \left(
\begin{array}{cc}
\left.{{\partial B_{n+2}}\over {\partial
B_n}}\right|_{\omega_c,x_c} & \left.{{\partial B_{n+2}}\over
{\partial X_n}}\right|_{\omega_c,x_c}\\ \left.{{\partial
X_{n+2}}\over {\partial B_n}}\right|_{\omega_c,x_c} &
\left.{{\partial X_{n+2}}\over {\partial
X_n}}\right|_{\omega_c,x_c}
\end{array}
\right)=
\left(
\begin{array}{cc}
\lambda_B^2 & 0\\ a & \lambda_1^2
\end{array}
\right) \label{Eq:LAR}
\end{equation}
with
\begin{equation}
\fl \lambda_1=\left[\left.{{\partial X_{n+2}}\over{\partial
X_n}}\right|_{\omega_c,x_c}\right]^{1/2}=
\left[\left.{{\partial{\cal X}_1}\over{\partial
X_n}}\right|_{\omega_c,x_c}\cdot \left.{{\partial{\cal
X}_2}\over{\partial
X_n}}\right|_{\omega_c,x_c}\right]^{1/2}=1.7412\ldots
\label{Eq:LAlambda1}
\end{equation}
and $a= \left.{{\partial X_{n+2}}\over{\partial
B_n}}\right|_{\omega_c,x_c}$. If we put $\langle
N\rangle_n={\omega\over {X_n}}{{\partial X_n}\over{\partial
\omega}}$ and $\langle M\rangle_n={k\over {X_n}}{{\partial
X_n}\over{\partial k}}$, at the transition, the average number of
steps on the wall grows as $\langle M\rangle_n= \lambda_1^n$.
Thus, $\langle M\rangle_n\sim L_n^{y_1}$ with
$y_1=\ln\lambda_1/\ln 2= 0.8001\ldots$ On the basis of
\eref{Eq:LAR} and \eref{Eq:LAlambda1} one can also write:
\begin{equation}
\fl
\langle N\rangle_n=\lambda_1^n+a\lambda_1^2\
{{\lambda_B^{n-2}-\lambda_1^{n-2}}\over{\lambda_B^2-\lambda_1^2}}
\label{Eq:LANfinal}
\end{equation}
This relation shows that, as long as $\lambda_1\leq\lambda_B$, in
the limit of large $n$, $\langle N\rangle_n$ scales as
$\lambda_B^n$. In other terms $\langle N\rangle_n\sim L_n^y$ with
$y=y_B$. On the other hand, if $\lambda_1>\lambda_B$, $\langle
N\rangle_n\sim\lambda_1^n$ and $y=\ln \lambda_1/\ln 2=y_1$. In
this latter case, which, as we will see below, is never realized
for ${\cal L}_A$, one must find $y_1=d_w$ because $\langle
M\rangle_n \sim \langle N\rangle_n$, which again means $\phi=1$.
At the adsorption fixed point of \eref{Eq:LAevenrec} ($k=k_c$),
$\lambda_1<\lambda_B$: thus, there, the SAW length in the presence
of wall scales as in the bulk without wall. The previous results
allow to write $\langle M\rangle_n\sim \langle N\rangle_n^\phi$
with $\phi=y_1/y_B=0.6773\ldots$

For any $k > k_c$ the SAW's are adsorbed on the wall and the
scalings of both $\langle M\rangle_n$ and $\langle N\rangle_n$ are
controlled by a fixed point at infinity. In this case both
quantities are proportional to $L_n^{d_w}$. On the other hand, for
$k < k_c$, we are in the normal regime in which $\langle
N\rangle_n\sim L_n^{y_B}$. For SAW's on Euclidean lattice, in this
last regime, $\langle M\rangle_n$ saturates to a constant value.
To the contrary, the analysis of recursion \eref{Eq:LAevenrec}
around $X=0.1536\ldots$ shows that $\langle M\rangle_n \sim
L_n^{-1.0844\ldots}$. This unphysical behavior is due to the
pathological increase of coordination with $n\to\infty$ typical of
hierarchical lattices. The above discussion can of course be
adapted to the case of odd $n$ levels, with the same scaling
results.

The method just illustrated can be easily generalized to fractal
walls with a whole range of $d_w$. In fact, adsorption on walls
with different dimensions can be obtained by simply changing the
form of the recursion. In general, in place of
\eref{Eq:LAevenrec}, we can have $p$ and $q$ nested applications
of the generating functions ${\cal X}_1$ and ${\cal X}_2$,
respectively. For example, for $p=2$ and $q=3$ we could have
\begin{eqnarray}
\fl &X_{n+p+q}=\nonumber \\
\fl &{\cal X}_1({\cal B}({\cal B}({\cal B}({\cal B}(B_n)))),
{\cal X}_2({\cal B}({\cal B}({\cal B}(B_n))),
{\cal X}_2({\cal B}({\cal B}(B_n)),
{\cal X}_1({\cal B}(B_n),
{\cal X}_2(B_n,X_n)))))
\end{eqnarray}
which corresponds to applying successively ${\cal X}_2$, ${\cal
X}_1$, ${\cal X}_2$, ${\cal X}_2$, and ${\cal X}_1$. The
corresponding wall has $d_w={{(p\ln 2+q\ln 3)}\over{(p+q)\ln 2}}$.
For given $p$ and $q$, we could generate many different recursions
by changing the order in which ${\cal X}_1$ and ${\cal X}_2$ are
applied. While this order influences the position of the
adsorption transition point, the corresponding critical exponents
are only functions of the wall dimension. This property follows
from the circumstance that by construction the different RG
recursions are diffeomorphically related. This is the mechanism by
which $d_w$ alone, and no other features of the wall, determines
the critical exponents of the transition. In these more general
conditions we find
\begin{equation}
\fl \lambda_1=\left[\left.{{\partial X_{n+p+q}}\over{\partial
X_n}}\right|_{\omega_c,x_c}\right]^{1/(p+q)}=
\left[\left(\left.{{\partial{\cal X}_1}\over{\partial
X}}\right|_{\omega_c,x_c}\right)^p\cdot
\left(\left.{{\partial{\cal X}_2}\over{\partial
X}}\right|_{\omega_c,x_c}\right)^q\right]^{1/(p+q)}
\label{Eq:LAgenericlambda1}
\end{equation}
where $x_c$ has to be calculated for the specific form of
$X_{n+p+q}$. As already mentioned, $d_w$ can be varied between $1$
and $\ln 3/\ln 2$ in ${\cal L}_A$. The lower limit corresponds to
$p=1$ and $q=0$. In this case we have $k_c=1$ and $\lambda_1=1$
which implies $\phi=0$. In fact, for this case of ``flat''
boundary, the adsorption transition fixed point merges with the
desorbed regime fixed point and becomes marginally unstable. This
is a peculiar feature due to the relatively too simple structure
of ${\cal L}_A$. On the other hand, the upper limit corresponds to
$p=0$ and $q=1$ for which one has $k_c=\sqrt{3+\sqrt{3}}$ and
$\lambda_1\equiv \lambda_B$. In this case we have exactly
$\phi=1$. For intermediate wall dimensions, $\phi$ is a monotonic
increasing function of $d_w$. A summary of our results for ${\cal
L}_A$ is reported in \Tref{Table:LA}.

$\phi<1$ implies continuity of the adsorption transition. On the
contrary, for $\phi=1$, at the transition $\langle
M\rangle_n\propto\langle N\rangle_n$, as for an adsorbed polymer.
In fact $\lambda_1\geq\lambda_B$ implies a discontinuity of
$\lim_{n\to\infty} [\langle M\rangle_n/\langle N\rangle_n]$ at the
adsorption point, i.e. a first--order transition. This
discontinuity, found only for $d_w=\ln 3/\ln 2$ in ${\cal L}_A$,
anticipates a more general result, valid for the other lattices we
considered: we find below that a sufficiently high $d_w$ can drive
polymer adsorption first--order. For ${\cal L}_A$ the threshold
condition for discontinuous adsorption occurs precisely when $d_w$
is at its maximum possible value. In general we will denote by
$d^*_w$ the value of $d_w$ above which $\phi=1$.

\begin{figure}
\vspace{6.0cm}
\includegraphics{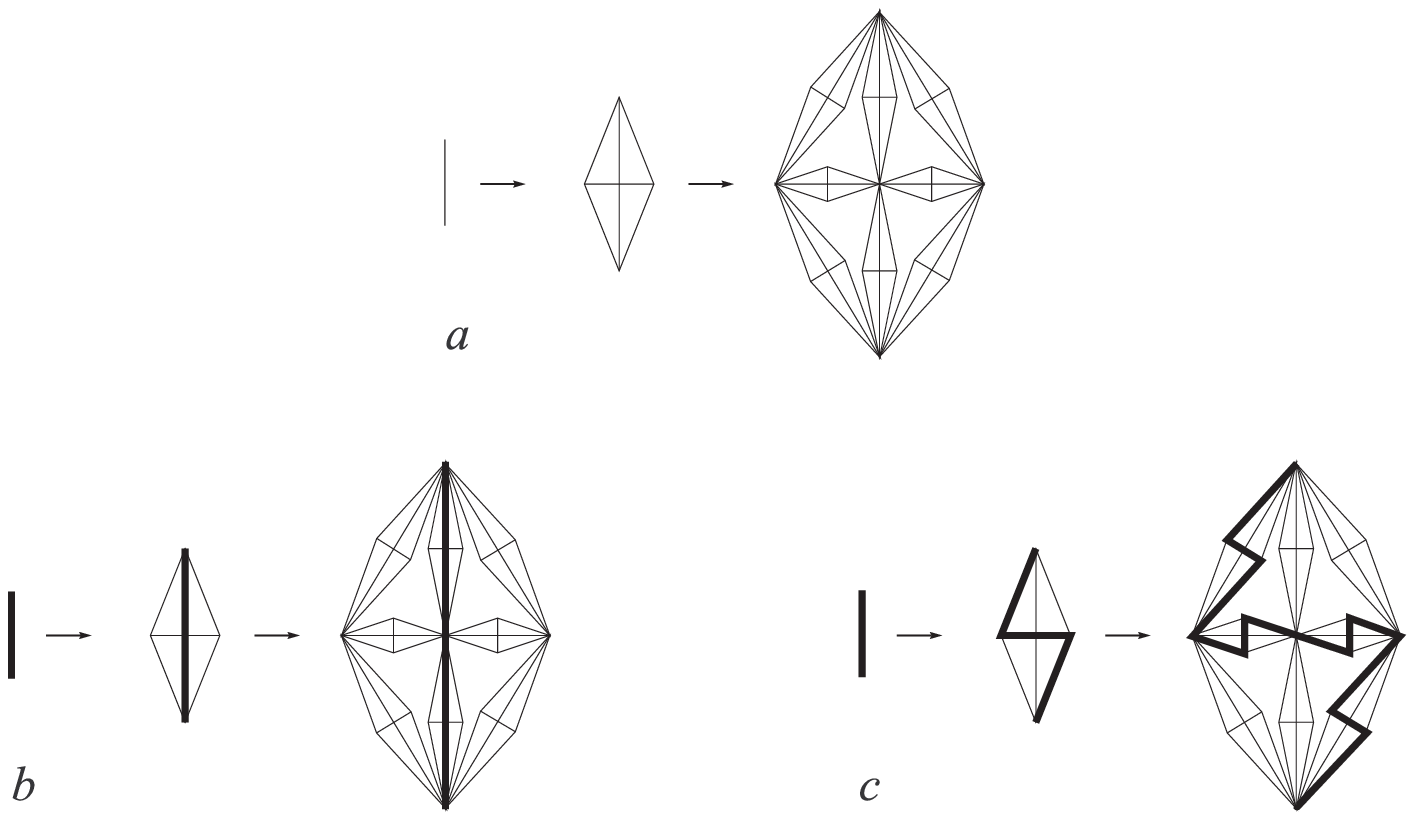}
\caption{\label{Fig:LB} $a$ Construction rule of ${\cal L}_B$.
$b$--$c$ Construction rules of deterministic fractal walls (heavy
lines) in ${\cal L}_B$.}
\end{figure}

\begin{figure}
\vspace{10.0cm}
\includegraphics{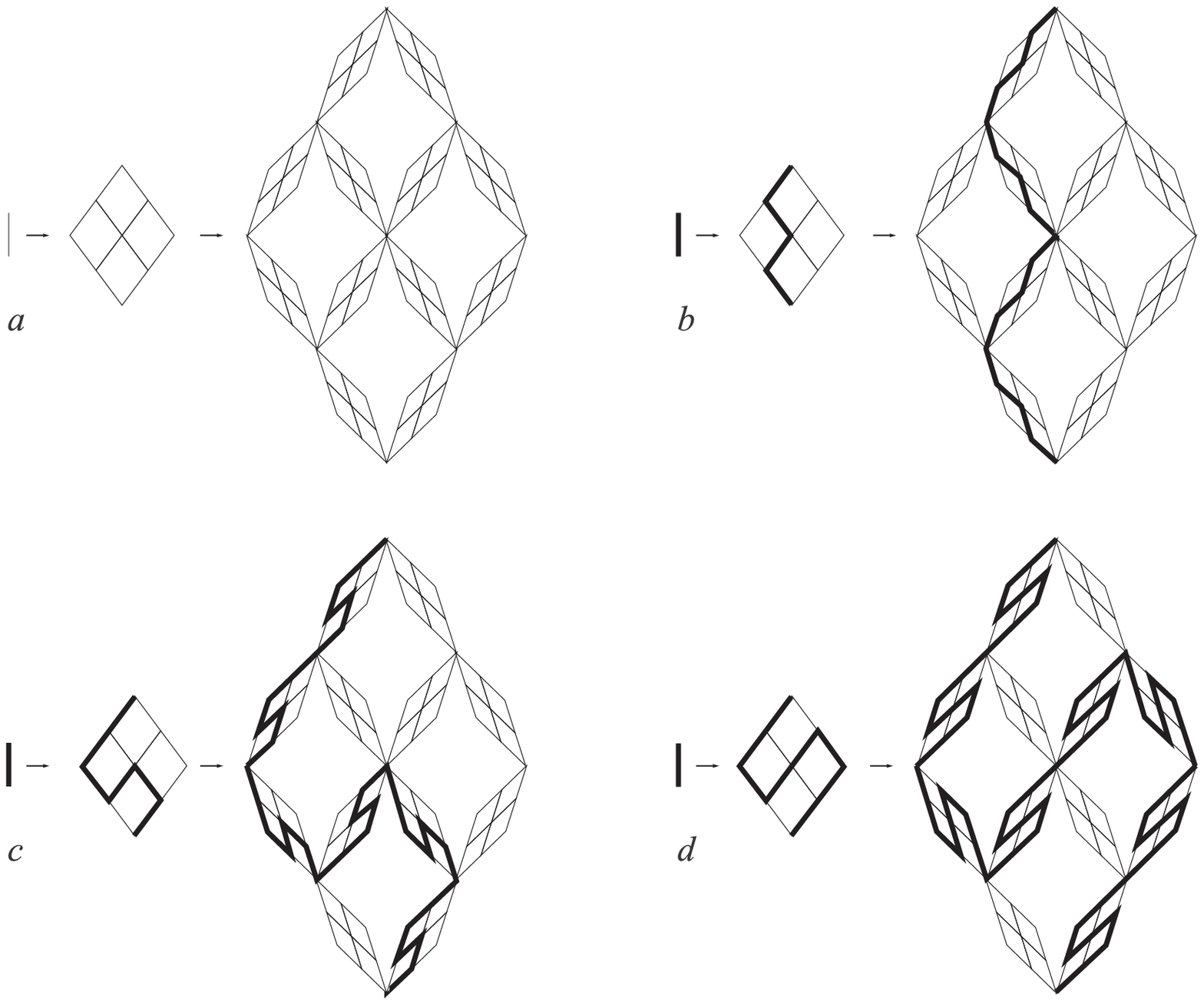}
\caption{\label{Fig:LC} $a$ Construction rule of ${\cal L}_C$.
$b$--$d$ Construction rules of deterministic fractal walls (heavy
lines) in ${\cal L}_C$.}
\end{figure}

To show that the scenario described above is not just peculiar of
${\cal L}_A$ and to investigate further the change of transition
order, we considered SAW adsorption on fractal walls also with
 ${\cal L}_B$ and ${\cal L}_C$,
sketched in Figures \ref{Fig:LB}a and \ref{Fig:LC}a, respectively.
${\cal L}_B$ has a diamond structure similar to that of ${\cal
L}_A$, but with a higher ramification and $d_L=3$. The bulk SAW
partition function obeys the recursion $B_{n+1}=3 B_n^2+4 B_n^3+2
B_n^4$ with $\omega_c=B^*=0.2441\ldots$ and $d_{saw}=1.1995\ldots$
In \Fref{Fig:LB}b and \ref{Fig:LB}c we report two construction
rules generating walls with dimensions $d_w^{(1)}=1$ and
$d_w^{(2)}=2$, respectively. The corresponding generating
functions are:
\begin{eqnarray}
\fl {\cal X}_1(X,B) &\equiv &X^2+2 X B^2+B^2 \label{Eq:LBgen1}\\
\fl {\cal X}_2(X,B) &\equiv &X B+X^2 B+X^4. \label{Eq:LBgen2}
\end{eqnarray}
For fractal walls obtained by suitably alternating the two rules,
\Tref{Table:LB} shows again that $\phi$ monotonically increases
with $d_w$. However, now for any $d_w > 1.67$ we have $\phi=1$.
Thus, when $d_w>d_w^*\simeq 1.67$ adsorption becomes
discontinuous.

Finally, we consider ${\cal L}_C$ for which $d_L=\ln 12/\ln
4=1.792\ldots$ \cite{LDM}. The bulk recursion is now $B_{n+1}=6
B_n^4+4 B_n^6+2 B_n^8$ with critical fixed point at
$\omega_c=B^*=0.5175\ldots$ and $d_{saw}=1.0649\ldots$
Unfortunately, neither $\omega_c$, nor $d_{saw}$ are too close to
the values appropriate for $2D$ square lattice, $0.378\ldots$ and
4/3 \cite{Nienhuis,V}, respectively. This occurs in spite of the
fact that ${\cal L}_C$ seems to mimic well the square lattice
structure at local level. ${\cal L}_C$ offers more possibilities
of wall construction rules. In Figures \ref{Fig:LC}b,
\ref{Fig:LC}c and \ref{Fig:LC}d we show three examples with
respective generating functions
\begin{eqnarray}
\fl {\cal X}_1(X,B) &\equiv &2 B^4+X^4+2 X^2 B^2+2 X^2 B^4
\label{Eq:LCgen1}\\
\fl {\cal X}_2(X,B) &\equiv &X B^3+X^2 B^4+X^5
B^3+X^2 B^2+X^3 B+X^6 \label{Eq:LCgen2}\\
\fl {\cal X}_3(X,B)
&\equiv &X^3 B+X^5 B+X^8\label{Eq:LCgen3}
\end{eqnarray}
The corresponding walls have $d_w^{(1)}=1$, $d_w^{(2)}=\ln 6/\ln
4= 1.292\ldots$ and $d_w^{(3)}=\ln 8/\ln 4=1.5$. Results for
${\cal L}_C$ are reported in \Tref{Table:LC}. For $d_w=1$
adsorption is continuous and $\phi=0.5437\ldots$ is not too far
from 1/2, the $\phi$ value for SAW adsorption on a smooth wall in
$2D$ \cite{BEG}. For increasing $d_w$, $\phi$ monotonically
increases and reaches a unit value at $d_w=d_w^*\simeq 1.4$. For
$d_w> d_w^*$, $\phi=1$ and the transition is always first--order.
Note that the first--order transitions found for $d_w>d_w^*$ in
Tables 2 and 3 correspond to $\lambda_1>\lambda_B$. From Tables
\ref{Table:LA}, \ref{Table:LB} and \ref{Table:LC} we also learn
that, upon increasing $d_w$, $k_c$ increases, as a rule, up to
small fluctuations caused by our peculiar recipe for varying
$d_w$. This indicates that increasing roughness makes adsorption
more difficult.

Our results suggest an interesting scenario for the adsorption
transition on fractal walls in more realistic models. First of
all, for a continuous adsorption transition, $\phi$ is an
increasing function of the wall dimension $d_w$. Moreover, for
high enough $d_w$, $\phi$ eventually reaches the value 1, its
upper bound marking the onset of first--order adsorption. This
fact is in open contrast with the results of Bouchaud and
Vannimenus \cite{BV} which, on the basis of scaling arguments,
suggested the bounds
\begin{equation}
\fl 1-{1\over{d_{saw}}}(d_L-d_w)\leq \phi \leq {{d_w}\over {d_L}}.
\label{Eq:phibounds}
\end{equation}
We find that $\phi$ does not satisfy \eref{Eq:phibounds}. Also the
lower bound in \eref{Eq:phibounds} is manifestly violated for low
enough $d_w$. Of course, here we deal with a hierarchical lattice,
which is not fully adequate to represent consistently all the
features of fractal objects. On the other hand, the bounds in
\eref{Eq:phibounds} were obtained by relying on formal analogies
with cases of regular geometry. Similar violations of these bounds
were reported previously, for both flat and fractal boundaries
\cite{ZMS,MMZ}. Here we identify in the monotonicity of $\phi$ and
in the tendency of the transition to turn first--order as $d_w$
increases the physical reasons for the upper bound violation. In
the investigation of Ref. \cite{MZM} first--order adsorption was
found for a particular choice of lattice and fractal boundary
among the many considered. A posteriori, we can understand that
result as due to the fact that such choice determines a rather
high $d_w$ relative to $d_{saw}$.

We obtained also results for random fractal walls. Unlike the case
of deterministic walls above, this problem cannot be solved
exactly, even on hierarchical lattices. However, quantities like
$\overline{\langle M\rangle}_n$ and $\overline{\langle
N\rangle}_n$, which now have to be averaged also over wall
randomness (overbar represents this average), can be calculated
quite accurately also for very large system sizes by a Monte Carlo
approach \cite{GS,SS,S}. We performed these calculations for
${\cal L}_C$ with walls obtained by random combined applications
of the rules in Figure 3c and 3d. At any level of lattice
construction, we choose whether the wall is realized by rule $3c$
or $3d$ with probabilities $1-\Delta$ and $\Delta$ ($\Delta<1$),
respectively. Of course, this determines which of the two
generating functions, ${\cal X}_2$ (\Eref{Eq:LCgen2}) and ${\cal
X}_3$ (\Eref{Eq:LCgen3}), has to be used in order to calculate
$X_{n+1}$ in terms of $X_n$. For $0<\Delta<1$ the wall has, on
average, a fractal dimension $d_w=\ln(6+2\Delta)/\ln4$. $X_n$
becomes now a random variable and we must consider its probability
distribution. Of course, we can only produce a finite sampling of
this distribution, by proceeding as follows. From a large set (up
to $4\cdot 10^5$ elements) $\{X_n\}$ of $n$--th level partition
values we generate each element of the new sample $\{X_{n+1}\}$ by
choosing first, with the appropriate probabilities, between rules
\eref{Eq:LCgen2} and \eref{Eq:LCgen3}; then from $\{X_n\}$ are
extracted at random the elements needed as entries into
\eref{Eq:LCgen2} or \eref{Eq:LCgen3}, and an element $X_{n+1}$ of
the new sample is computed as a function of them and of $B_n$.
Using some numerical tricks to control the possible rapid
divergence of the partition functions near the transition, we
could iterate this procedure up to $n=30\div 35$.

By analyzing the scaling of $\overline{\langle M\rangle}_n$ and
$\overline{\langle N\rangle}_n$ at the transition point (which now
has to be numerically determined) as a function of $L_n$, we could
estimate $\phi$ for different $\Delta$'s as reported in Figure 4.
Even if the relative poorness of the samplings causes appreciable
uncertainty in the $\phi$ determinations (Figure 4), we see that
this exponent stabilizes for $\Delta > 1/2$ to a value
compatible with the upper limit $\phi=1$. The results are
consistent with the scenario for deterministic walls. In
particular, $\Delta= 1/2$ corresponds to $d_w= 1.404$, extremely
close to $d_w^*\simeq 1.4$ applying in that case. All this
suggests that first--order adsorption should be expected also with
random fractal walls, with the same threshold $d_w^*$ as in the
deterministic case.

\section{Conclusions}

Our most remarkable result here is the roughness induced change
into first--order of the adsorption transition.  Without any other
changes in the polymer--wall interactions, high enough $d_w$ make
the adsorption transition discontinuous. This is found in all
lattices considered, although marginally in ${\cal L}_A$. The
change in the nature of the transition takes place for $d_w$
definitely larger than $d_{saw}$. In spite of the qualitative
value of our model calculations, we can hope that similar
properties could hold in realistic situations, in both $2D$ and
$3D$.

The change into first-order of the adsorption transition should be
imputed to the fact that, upon increasing $d_w$, the drop in
entropy associated to a localization of the polymer near the wall,
increases with $d_w$. Even if it is not easy to give a precise
meaning to the notion of distance of a polymer from a fractal
substrate, at qualitative level we can argue that the entropic
repulsion effect due to localization \cite{deGennes}, should
create a free energy barrier, whose height and (long) range
certainly increase as $d_w$ gets larger. The rougher the surface,
the more it limits the configurations of the confined polymer.
Thus, we can think $d_w^*$ as the maximum dimension for which
``tunneling'' of the polymer can still occur continuously from the
attractive free energy well at small distance to the unbound state
at infinite distance across the barrier. For $d_w>d_w^*$, the
``tunneling'' becomes discontinuous because of the too large
barrier (this means that the SAW, right at the desorption point,
is not delocalized yet, as occurs in continuum adsorption).

A mechanism like the qualitative one outlined above has been
demonstrated and precisely described for the phenomenon of wetting
of self--affine rough substrates in $2D$, which has some analogies
with our adsorption \cite{SS}. Indeed, in that case one finds that
a fluctuating interface depinns discontinuously from a rough
substrate as soon as the roughness of the latter (measured by its
self--affinity exponent $\zeta_w$) exceed $\zeta_0$, the exponent
specifying the intrinsic roughness of the interface in the bulk
\cite{GS,SS}. The coincidence of the threshold roughness with
$\zeta_0$ (which by analogy, would suggest $d_w=d_w^*$ here) is a
peculiar feature of the interfacial problem in $2D$. Indeed, for
that problem, the possibility of a path--integral description in
$1+1$ dimensions allows to establish a correspondence with the
$1D$ quantum tunneling of a particle across a long range repulsive
potential barrier. It turns then out that a roughness
$\zeta_w=\zeta_0=1/2$ determines a decay of this potential right
at the threshold for discontinuous tunneling \cite{SS}. Our
results here show that fractal wall roughness leaves room for a
continuous polymer adsorption also when $d_w>d_{saw}$.

\begin{table}
\caption{SAW adsorption on deterministic fractal walls in ${\cal
L}_A$. Wall rules are applied in the reported order.}
\begin{indented}
\item[]\begin{tabular}{llll}
\hline \multicolumn{4}{c}{$d_L=\ln 5/\ln 2= 2.3219$,
$\omega_c=0.3660$, $d_{saw}=1.1814$}\\ \hline
 &$d_w$&$k_c$&$\phi$\\ \hline
${\cal X}_1$&1&1&0\\
${\cal X}_2,{\cal X}_1,{\cal X}_1,{\cal X}_1,{\cal X}_1,{\cal
X}_1$&1.0975&1.7398&0.3856\\
${\cal X}_2,{\cal X}_1,{\cal X}_1,{\cal X}_1,{\cal X}_1$&1.1170
&1.7725&0.4230\\
${\cal X}_2,{\cal X}_1,{\cal X}_1,{\cal X}_1$&1.1462&1.8165&0.4736\\
${\cal X}_2,{\cal X}_1,{\cal X}_1$&1.1950&1.8798&0.5483\\
${\cal X}_2,{\cal X}_2,{\cal X}_1,{\cal X}_1,{\cal X}_1$&1.2340&2.0267
&0.5957\\
${\cal X}_2,{\cal X}_1$&1.2925&1.9806&0.6773\\
${\cal X}_2,{\cal X}_2,{\cal X}_1$&1.3900&2.1051&0.7904\\
${\cal X}_2$&1.5850&2.1753&1$^a$\\
\hline
\end{tabular}
\item[$^a$] $\lambda_1=\lambda_B$
\end{indented}
\label{Table:LA}
\end{table}

\begin{table}
\caption{SAW adsorption on deterministic fractal walls in
${\cal L}_B$.}
\begin{indented}
\item[]\begin{tabular}{llll}
\hline \multicolumn{4}{c}{$d_L=3$, $\omega_c=0.2441$,
$d_{saw}=1.1995$}\\
\hline &$d_w$&$k_c$&$\phi$\\
\hline
${\cal X}_1$&1&3.3049&0.6613\\
${\cal X}_2,{\cal X}_1,{\cal X}_1,{\cal X}_1$ &5/4&3.3476&0.8124\\
${\cal X}_2,{\cal X}_1,{\cal X}_1$ &4/3&3.3500&0.8628\\
${\cal X}_2,{\cal X}_1$&3/2&3.3549&0.9637\\
${\cal X}_2,{\cal X}_2,{\cal X}_1$&5/3&3.3629&1$^a$\\
${\cal X}_2,{\cal X}_2,{\cal X}_2,{\cal X}_1$ &7/4&3.3652&1$^a$\\
${\cal X}_2$&2&3.3663&1$^a$\\
\hline
\end{tabular}
\item[$^a$] $\lambda_1>\lambda_B$
\end{indented}
\label{Table:LB}
\end{table}

\begin{table}
\caption{SAW adsorption on deterministic fractal walls in
${\cal L}_C$.}
\begin{indented}
\item[]\begin{tabular}{llll}
\hline \multicolumn{4}{c}{$d_L=\ln 12/\ln 4= 1.7925$,
$\omega_c=0.5175$, $d_{saw}=1.0649$}\\
\hline & $d_w$ &
$k_c$ & $\phi$ \\
\hline
${\cal X}_1$ & 1 & 1.3279 & 0.5437 \\
${\cal X}_2$ & 1.2925 & 1.4715 & 0.8140 \\
${\cal X}_3,{\cal X}_2$ & 1.3962 & 1.6287 & 0.9741 \\
${\cal X}_3,{\cal X}_3,{\cal X}_2$ & 1.4308 & 1.6514 & 1$^a$\\
${\cal X}_3$ & 1.5 & 1.6565 &1$^a$\\
\hline
\end{tabular}
\item[$^a$] $\lambda_1>\lambda_B$
\end{indented}
\label{Table:LC}
\end{table}

\newpage

\begin{figure}
\vspace{6.5cm} \includegraphics{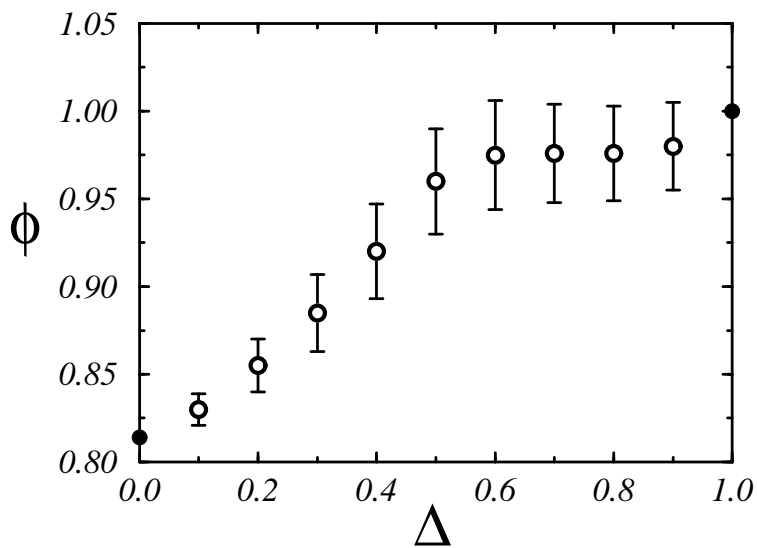} \caption{\label{Fig:LC_P0} $\phi$ as a
function of $\Delta$ for SAW adsorption on random fractal walls in
${\cal L}_C$. $\phi$ values at $\Delta=0$ and $\Delta=1$ (closed
circles) are exact.}
\end{figure}

\end{document}